\documentclass[aps,prc,twocolumn,floatfix,12pts,superscriptaddress]{revtex4}
\usepackage{epsfig}
\usepackage{amssymb}
\usepackage{amsmath}
\usepackage{color}
\usepackage{graphicx}
\usepackage{epsfig}
\usepackage{amsmath}
\usepackage{color}
\usepackage{graphicx}
\usepackage{float}
\usepackage{hyperref}
\definecolor{blue}{rgb}{0.05, 0.05, 0.5}
\def \beq{\begin{equation}}
\def \eeq{\end{equation}}
\def \beqa{\begin{eqnarray}}
\def \eeqa{\end{eqnarray}}

\usepackage{xspace}

\begin{document}
\title{Production and anisotropic flow of thermal photons in collision of $\alpha$-clustered carbon with heavy nuclei at relativistic energies}
\author{Pingal Dasgupta}
\affiliation{Key Laboratory of Nuclear Physics and Ion-beam Application (MOE), Institute of Modern Physics, Fudan University, Shanghai 200433, China}
\affiliation{Shanghai Research Center for Theoretical Nuclear Physics, NSFC and Fudan University, Shanghai $200438$, China}
\author{Rupa Chatterjee}
\email[]{rupa@vecc.gov.in}
\affiliation{Variable Energy Cyclotron Centre, 1/AF, Bidhan Nagar, Kolkata-700064, India}
\affiliation{Homi Bhabha National Institute, Training School Complex, Anushaktinagar, Mumbai 400094, India}
\author{Guo-Liang Ma}
\email[]{glma@fudan.edu.cn}
\affiliation{Key Laboratory of Nuclear Physics and Ion-beam Application (MOE), Institute of Modern Physics, Fudan University, Shanghai 200433, China}
\affiliation{Shanghai Research Center for Theoretical Nuclear Physics, NSFC and Fudan University, Shanghai $200438$, China}

\begin{abstract}
The presence of $\alpha$-clustered structure in the light nuclei produces different exotic shapes in nuclear structure studies at low energies.  Recent phenomenological studies suggest that collision of heavy nuclei with $\alpha$-clustered carbon ($^{12}$C) at relativistic energies can lead to large initial state anisotropies. This is expected to impact the final momentum anisotropies of the produced particles significantly. The emission of electromagnetic radiations is considered to be more sensitive to the initial state compared to hadronic observables and thus photon observables are expected to be affected by the initial clustered structure profoundly.  In this work we estimate the production and anisotropic flow of photons from most-central collisions of triangular $\alpha$-clustered carbon and gold at $\sqrt{s_{\rm NN}}=200$ GeV using an event-by-event hydrodynamic framework and compare the results with those obtained from unclustered carbon and gold collisions. We show that the thermal photon $v_3$ for most central collisions is significantly large for the clustered case compared to the case with unclustered carbon, whereas the elliptic flow parameter does not show much difference for the two cases. In addition, the ratio of anisotropic flow coefficients is found to be a potential observable to constrain the initial state produced in relativistic heavy-ion collisions and also to know more about the $\alpha$-clustered structure in carbon nucleus.

\end{abstract}

\maketitle

\section{Introduction}
Direct photons are considered as one of the cleanest probes to study the initial state and the evolution of the hot and dense matter produced in relativistic heavy-ion collisions. Various properties of the direct photon spectra and anisotropic flow parameters have been explored in detail in past two decades~\cite{Srivastava:2008es, Chatterjee:2005de, Gale:2014dfa, Chatterjee:2013naa, Monnai:2014kqa, McLerran:2014hza, Basar:2012bp, Tuchin:2012mf, Zakharov:2016mmc, Chatterjee:2021gwa}. The most salient feature of the photon observables is their strong sensitivity to the initial conditions. Different collision geometries, initial-state fluctuations, the inclusion of initial state nucleon shadowing as well as slight variation of initial parameters in the model calculation have been found to affect the anisotropic flow parameters of photons significantly~\cite{Dasgupta:2020orj, Dasgupta:2016qkq, Vujanovic:2014xva, Liu:2012ax, Chatterjee:2008tp, Chatterjee:2009qz, Chatterjee:2012dn}. Thus, the study of direct photons from relativistic nuclear collisions offers an excellent opportunity to explore the hot and dense initial stage of Quark-Gluon Plasma (QGP) produced in those collisions. However, it is important to mention that the inconsistency between the experimental photon anisotropic flow data and the results from theory calculations has been a subject of research for quite some time~\cite{Adare:2015lcd,Acharya:2018bdy}. Realistic corrections in the initial conditions, upgraded hydrodynamical framework, and modified rates of thermal photon production have been found to improve the theoretical calculation significantly~\cite{David:2019wpt, Shen:2015nto, Gale:2020xlg, Dasgupta:2017fns, Iatrakis:2016ugz}.

In recent times, it has been shown that a large anisotropic flow of charged hadrons can appear even in small collision systems such as ${\rm p + p}$, ${\rm p + A}$, ${\rm d + A}$, ${\rm He + A}$ etc.~\cite{PHENIX:2018lia, ATLAS:2015hzw, CMS:2012qk}. Significant initial-state anisotropies play a pivotal role in building up large final-state anisotropies for small systems. Recent interesting studies have suggested that the geometric effects of $\alpha$-clustering in the light nuclei ({$^{7,9}$Be}, $^{12}$C, $^{16}$O, etc.) can also be realized in the realm of relativistic nuclear collisions~\cite{Rybczynski:2017nrx,Bozek:2014cva,Li:2020vrg,Zhang:2017xda,He:2021uko}. { Similar attempts to identify nuclear deformations in heavy nuclei can be found in several studies~\cite{Giacalone:2021udy,Jia:2021tzt,Jia:2021qyu,Xu:2021vpn,Zhao:2022grq,Bally:2022vgo}.}

Clustering has long been known to play a key role in understanding the structure of light nuclei. Over the past half century, complex clusters of light nuclei have been discovered, especially for the typical 3$\alpha$-clustered and 4$\alpha$-clustered structures in $^{12}$C and $^{16}$O~\cite{gamow, Cook:1957zz,Hoyle:1954zz, Freer:2017gip,Liu:2018eyg,Ren:2018xpt}. Many theories, such as AMD~\cite{Kanada-Enyo:2006rjf}, FMD~\cite{Neff:2003ib}, THSR~\cite{Tohsaki:2001an,Zhou:2016mhb}, and effective field-theory lattice calculations~\cite{Lahde:2013png} attempt to explore possible geometries and even non-rigid structures of light nuclei. At present, the theoretical elucidation of the $\alpha$-structure in $^{12}$C and $^{16}$O is still a hard problem. Therefore, it is desirable to study the problem with some experimental probes~\cite{He:2014iqa,He:2016cwt,Huang:2017ysr,Shi:2021far,Guo:2017tco}.
Thus, experiments incorporating carbon and oxygen at the relativistic colliders can be useful to shed light on the clustered structures. Studies based on kinetic theory and hydrodynamic model simulations have shown definitive and significant effects of $\alpha$-clustering on the anisotropic flow of hadrons   at different collision energies lately~\cite{Rybczynski:2017nrx,Bozek:2014cva,Li:2020vrg,Zhang:2017xda,He:2021uko}. The effect of $\alpha$-clustering on the direct photon signals in ${\rm C+Au}$ collisions at $200A$ GeV was first studied in Ref.~\cite{Dasgupta:2020orj}. It has been shown that specific orientations of the $\alpha$-clustered C+Au collisions give rise to larger triangular flow ($v_3$) or elliptic flow ($v_2$) parameters considering a smooth initial energy density distribution~\cite{Dasgupta:2020orj}. In this work, we use a more realistic hydrodynamic framework to study the effect of clustered structure on different photon anisotropic flow  ($v_1,v_2$ and $v_3$) observables in the most-central collision scenario. { We expect that the properties emerging from this study will be useful for understanding the behavior of photon anisotropic flow parameters from similar colliding systems with $\alpha$-clustered carbon at different beam energies.}

The paper is organized as follows. In the next section, we briefly discuss the initial parameters and the framework for the model calculation. In section III,  we discuss the results of thermal photon spectra and anisotropic flow coefficients, and finally, we summarize the results in section IV.

\section{Framework}
{In this study, we adapt a similar procedure used in Refs.~\cite{Bozek:2014cva,Dasgupta:2020orj} to prepare the initial conditions for $\alpha$-clustered C+Au collisions at $200A$ GeV  (see Appendix-I).  The initial conditions are subsequently evolved with a (2+1) dimensional longitudinally boost invariant ideal relativistic hydrodynamic framework~\cite{Holopainen:2010gz}  to obtain the space-time evolution at mid-rapidity. The value of initial thermalization time is taken as $\tau_0=$ 0.17 fm/$c$  and the transverse components of initial flow velocities (i.e., $v_x$ and $v_y$) are neglected.  A lattice-based equation of state {(L\&S)}~\cite{Laine:2006cp} is used in the hydrodynamic model and the constant freeze-out temperature ($ T_f$) is taken as 160 MeV.

The thermal photon production from individual events  is estimated by integrating the  emission rates (i.e., $R=E\frac{dN}{d^3pd^4x}$) over the entire space-time evolution:
		\begin{eqnarray}
	{E \frac{dN}{d^3p} \ =  \int \it{R}\big(E^*(x),T(x)\big)d^4x}
	\label{dn_phot}
\end{eqnarray}
The $T(x)$ in the above equation is the local temperature and $E^*(x)=p^\mu u_\mu(x)$, where $p^\mu$  represents the four-momentum of photon and $u_\mu$ is the local 4-velocity of the flow field. We use the complete next-to-leading order emission rates from Refs.~\cite{Arnold:2001ms,Ghiglieri:2013gia}  to evaluate photon production from the  QGP phase and the parameterized rates from Ref. ~\cite{Turbide:2003si}  for the hadronic sector. \\
The differential anisotropic flow coefficients ($v_n(p_T)$) and initial-state eccentricities are evaluated using the equations shown in Appendix-I. }

\section{Production and anisotropic flow of thermal photons}

In the present study, we focus on the most central collisions where more geometry-dominated effects are expected to be seen. As presently we do not have any knowledge about the minimum bias event distribution of charged hadrons from hydrodynamic simulations, we have chosen events with  $N_{\rm part} >80$ for both the clustered and unclustered cases to understand the effects in the most-central collision scenario. It should be noted that such an event selection criterion in the Glauber model calculations with GLISSANDO ~\cite{Broniowski:2007nz} refers to almost similar centrality classes ($\approx 0-1\%$) for both the clustered and unclustered cases. To study the effect of initial clustered structure on photon observables, we have considered sufficiently large number of random events for both clustered and unclustered cases with the event selection criterion  $ N_{\rm part} >80$.  \\

\begin{figure}[tb]
	\centering
	{\includegraphics*[scale=0.4,clip=true]{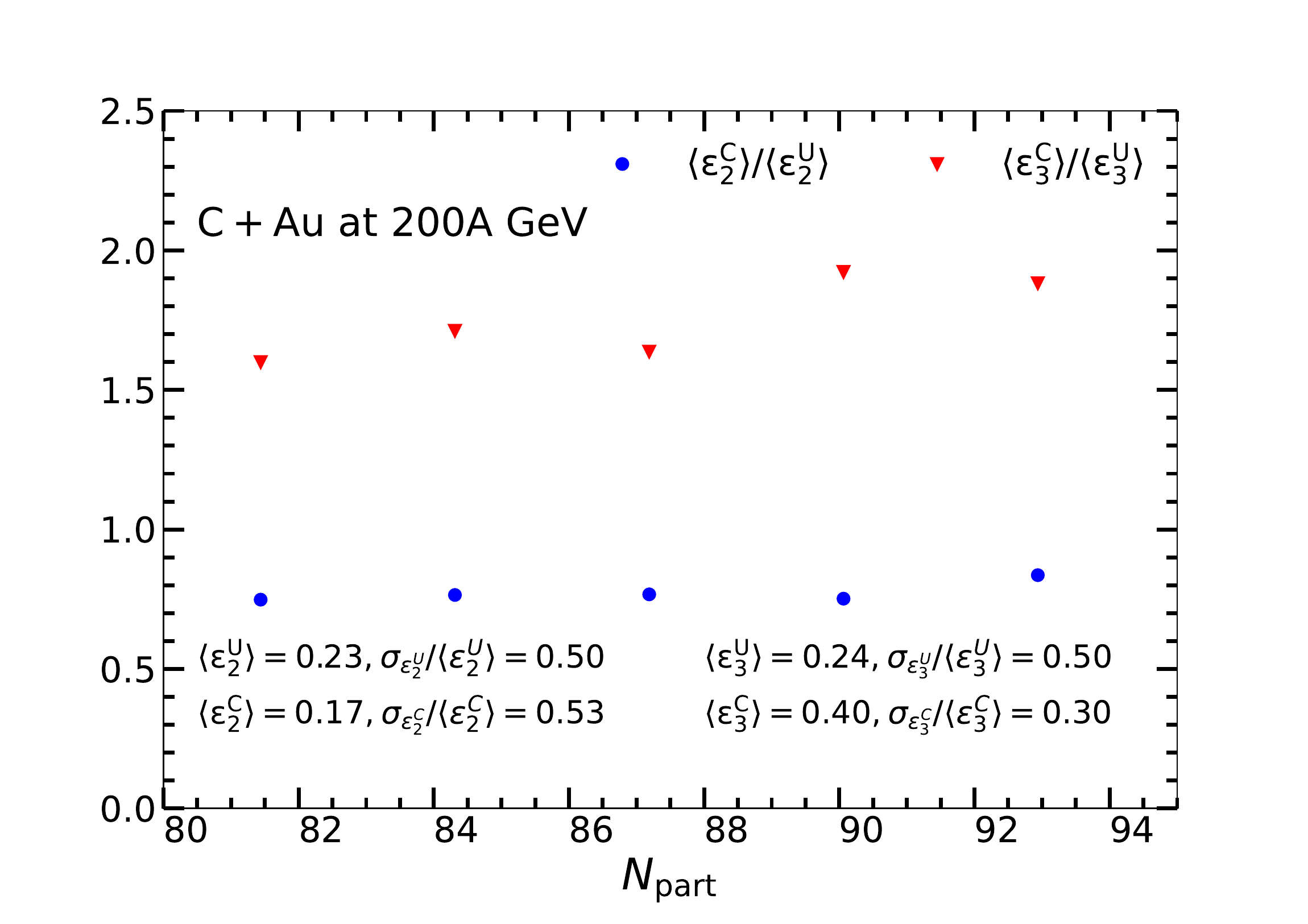}}
	\vspace{-3mm}
	\caption {(Color online) The ratio of $\alpha$-clustered to unclustered event-averaged initial state elliptic and triangular eccentricities as a function of $N_{\rm part}$ from  C+Au collisions at $200A$ GeV. }
	\label{fig1}
\end{figure}

 In Fig.~\ref{fig1}, we show the $ N_{\rm part}$ dependent behavior of clustered to unclustered ratio of the event-averaged (over 500000 events) initial elliptic  ($ \langle \varepsilon_2^C\rangle/\langle \varepsilon_2^U\rangle$)  and triangular eccentricities ($ \langle \varepsilon_3^C\rangle/\langle \varepsilon_3^U\rangle$). We find that the ratio of elliptic eccentricities at various $ N_{\rm part}$ are slightly less than 1, whereas,  the  the ratio for triangular eccentricities varies in the range $1.6-2.0$ as a function of $N_{\rm part}$.  This clearly shows that the triangular anisotropies for the clustered case show more than 50\% rise than the unclustered case for most central collisions. {We have checked the initial eccentricities also with the TRENTO initial condition~\cite{Moreland:2014oya} and found a similar behavior of elliptic and triangular eccentricities with $N_{\rm part}$ for both the cluster and uncluster cases.} \\

\begin{figure}[htbp!]
	\centering
	{\includegraphics*[scale=0.4,clip=true]{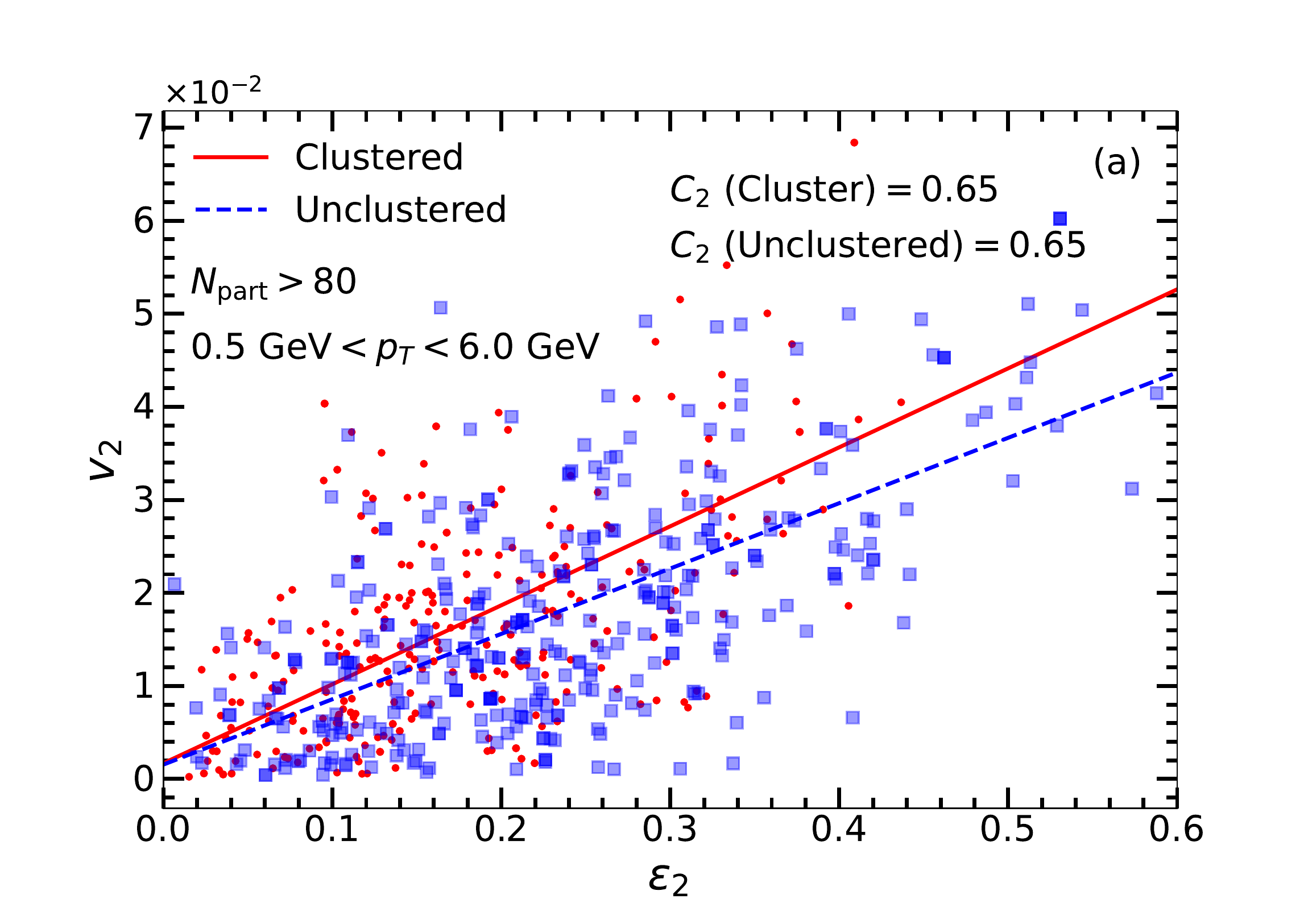}}
	{\includegraphics*[scale=0.4,clip=true]{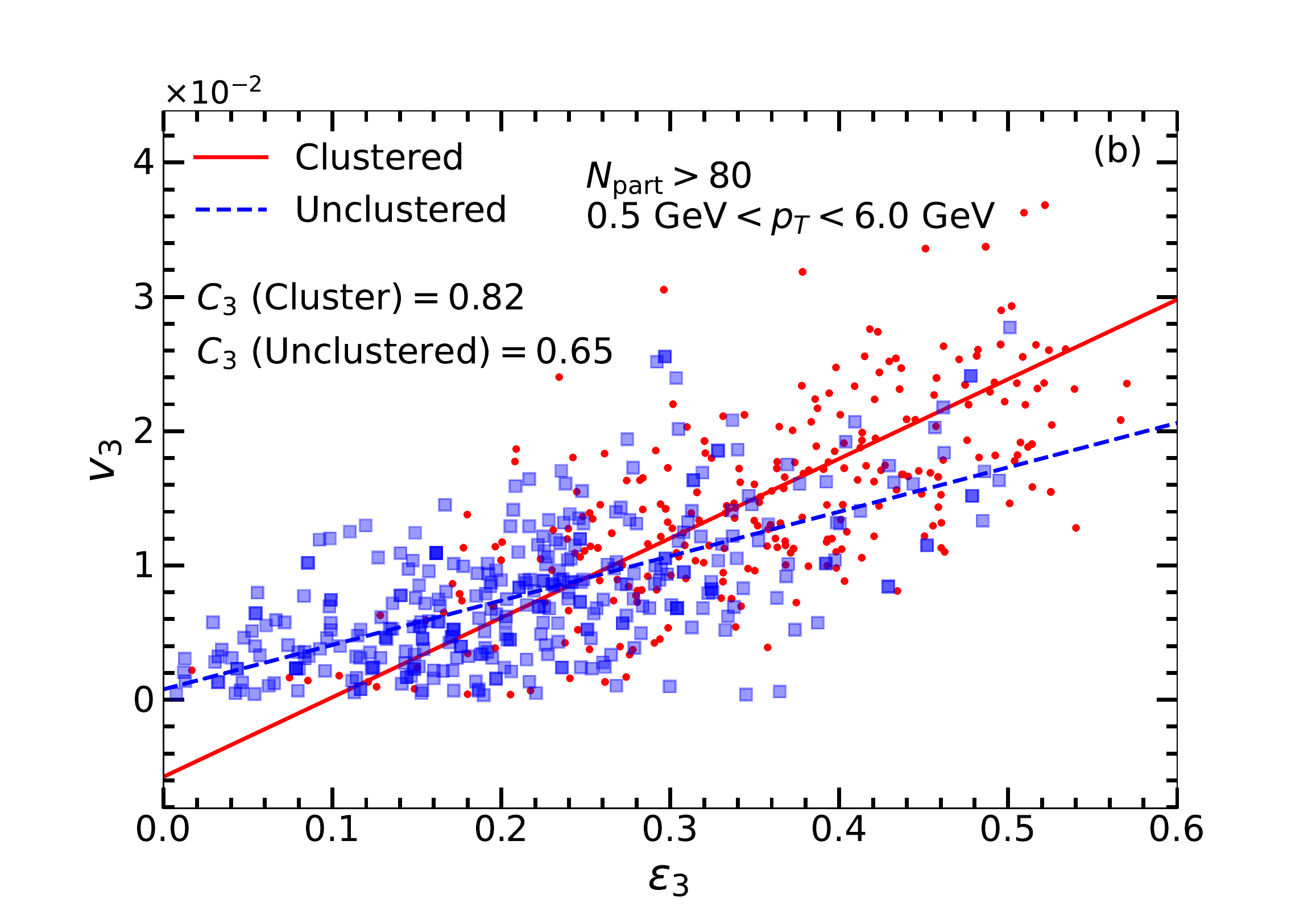}}
	\vspace{-3mm}
	\caption {(Color online)  The linear correlation between  (a)  $v_2$--$\varepsilon_2$ and (b)  $v_3$--$\varepsilon_3$  for  the $\alpha$-clustered C+Au and the unclustered C+Au collisions at $200A$ GeV. }
	\label{fig2}
\end{figure} 

The event distribution of the $\alpha$-clustered C+Au and the unclustered C+Au collisions in the $v_n$ -- $\varepsilon_n$ plane is shown in  Fig.~\ref{fig2}. The integrated $v_n$ is obtained by integrating the differential thermal photon flow over the  $p_T$ range $0.5 - 6.0$ GeV. The linear correlation  (Pearson correlation) coefficient $C_n(\varepsilon_n,v_n)$ for both clustered and unclustered cases is shown for comparison.  The correlation coefficient is defined as:
\begin{eqnarray}
C_n(\varepsilon_n,v_n)=\bigg\langle \frac{(\varepsilon_n-\langle \varepsilon_n\rangle)(v_n-\langle v_n\rangle)}{\sigma_{\varepsilon_n} \sigma_{v_n}}\bigg\rangle,
\end{eqnarray}
where the quantities without (with) angular bracket denote single event (event-averaged) values, $\sigma_{\varepsilon_n}$ and $\sigma_{v_n}$ are the standard deviations of $\varepsilon_n$ and $v_n$, respectively. To calculate the average of $v_n$, we use the integrated photon yield as weight, whereas, the total deposited energy on the transverse plane has been considered as the weight for calculating the average of initial state anisotropy. We see that the correlation coefficient between $v_2$ and $\varepsilon_2$ for the clustered and unclustered cases is similar (about 0.65).  On the other hand, we see a stronger linear correlation  (coefficient $\approx 0.82$) in $v_3$ -- $\varepsilon_3$ plane for the clustered carbon compared to the unclustered case (coefficient $\approx 0.65$). \\

  \begin{figure}[htbp!]
	\centerline{\includegraphics*[scale=0.38,clip=true]{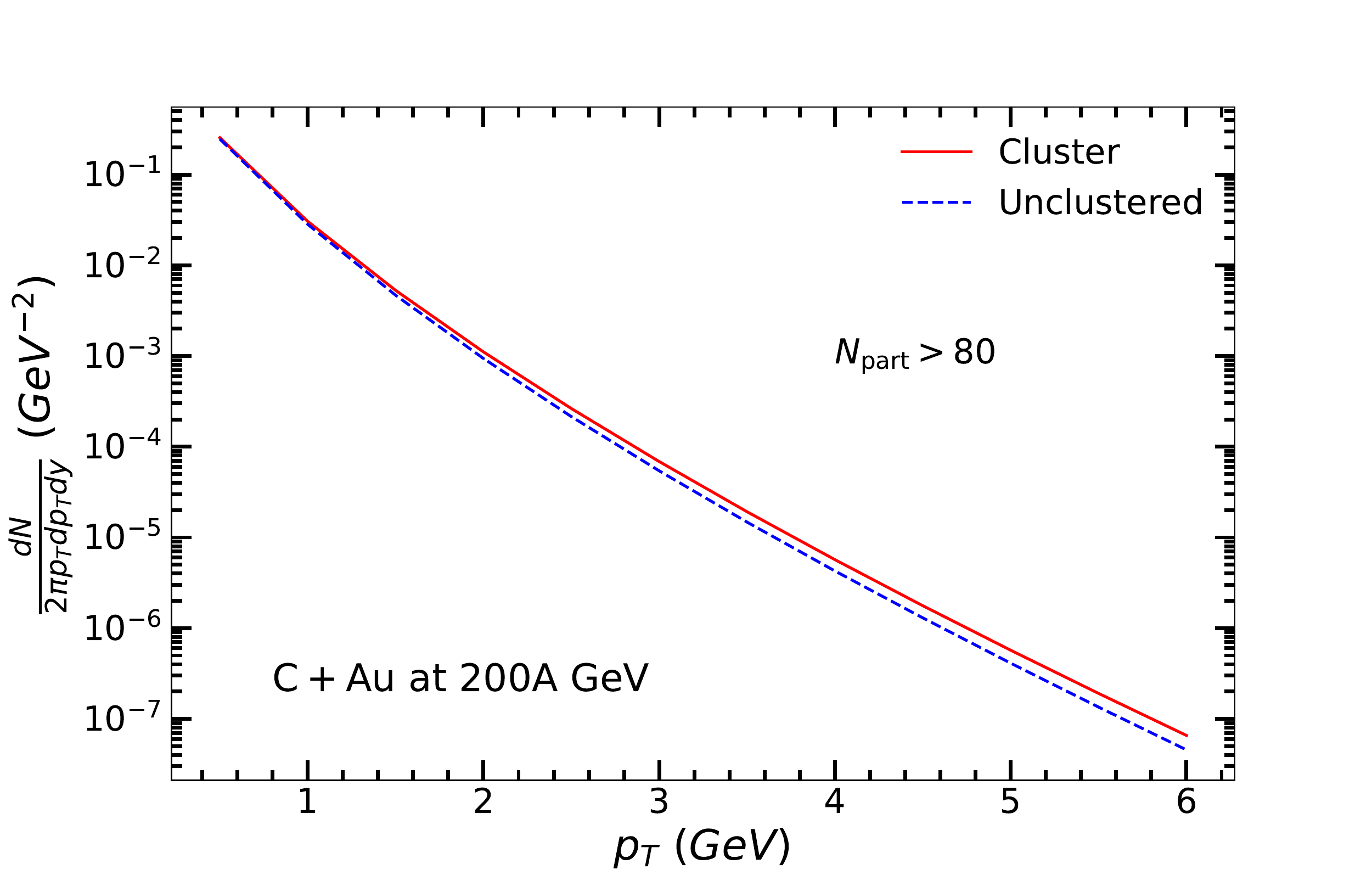}}
	\caption{(Color online) The event-averaged thermal photon spectra from $\alpha$-clustered C + Au and unclustered C + Au collisions at $200A$ GeV.}
	\label{fig3}
\end{figure}

In Fig.~\ref{fig3}, we show a comparison of the event-averaged thermal photon spectra from $\alpha$-clustered and unclustered C+Au collisions. The spectra for both the cases are found to be close to each other. In the region $ p_T > 4$ GeV, we observe a slight excess of photon production for the clustered case in comparison to the unclustered case which perhaps occurs due to the presence of initial hot spots in the clustered carbon case.\\

\begin{figure}[htbp!]
	\centering
	{\includegraphics*[scale=0.4,clip=true]{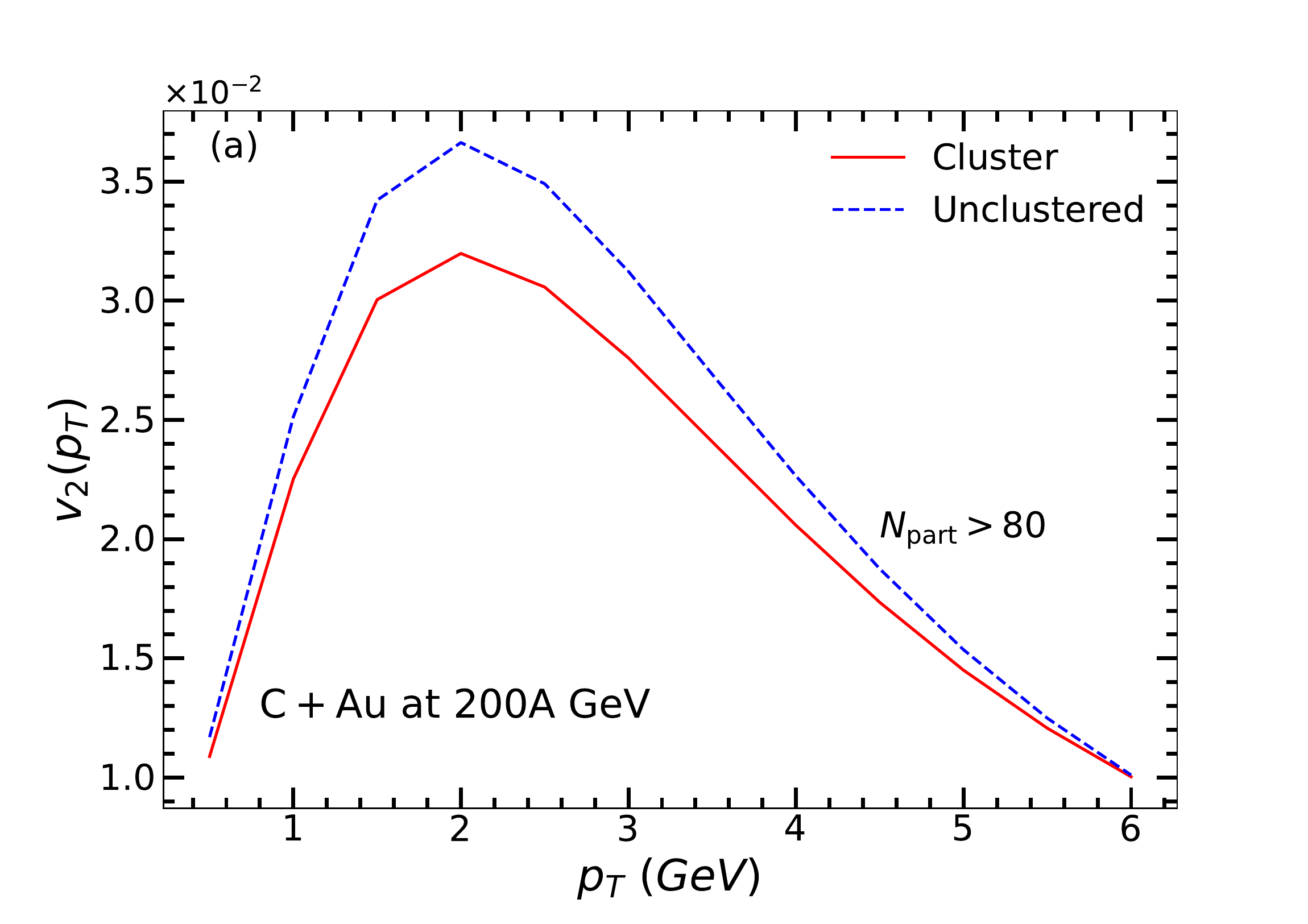}}
	{\includegraphics*[scale=0.4,clip=true]{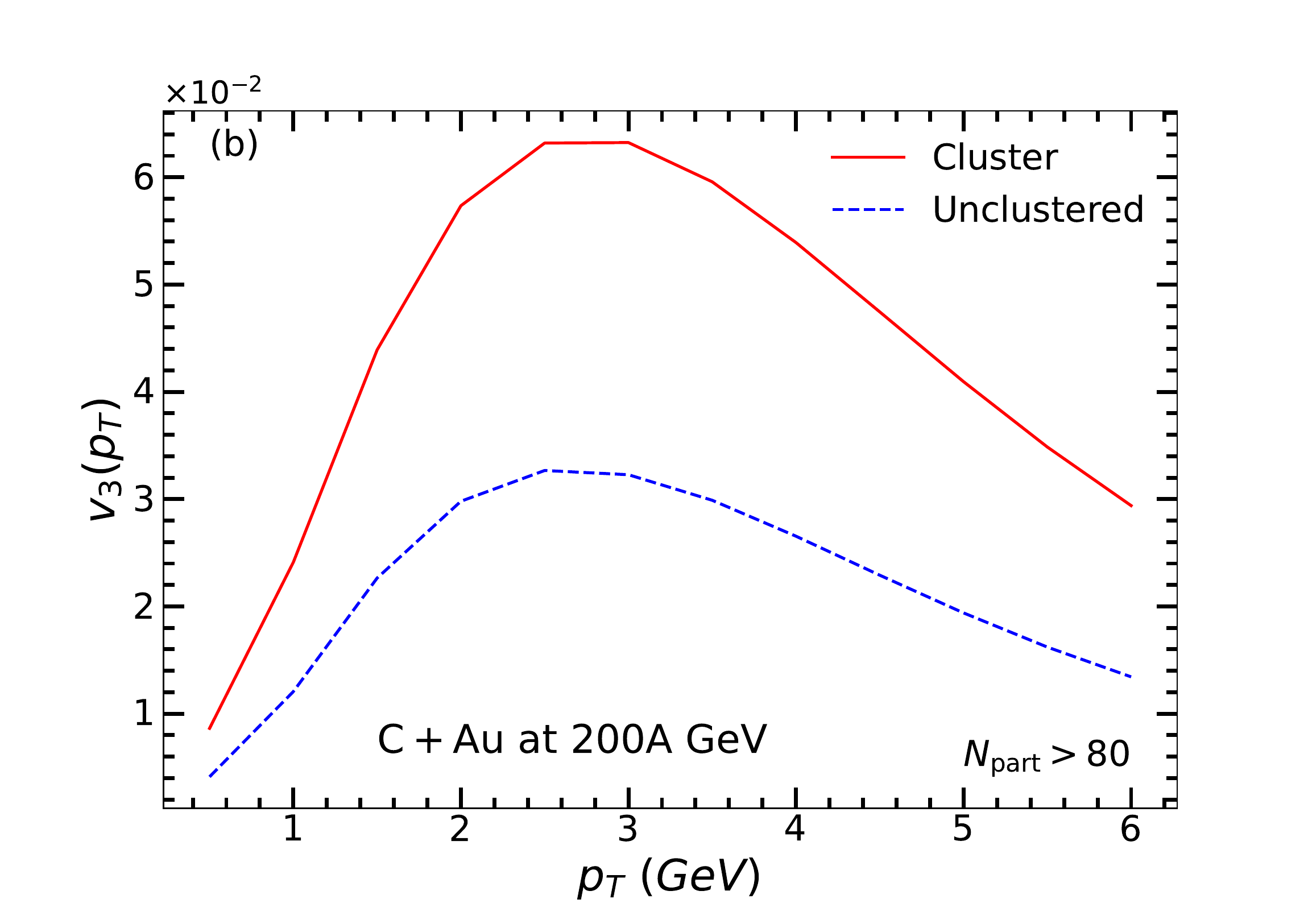}}
	\vspace{-3mm}
	\caption {(Color online) (a) Elliptic and (b) triangular flow of thermal photons as a function of $p_T$ from the  $\alpha$-clustered and unclustered C + Au collisions at $200A$ GeV.}
	\label{fig4}
\end{figure}

The event-averaged elliptic and triangular flow parameters of thermal photons as a function of $ p_T$ are presented in Fig.~\ref{fig4}(a) and Fig.~\ref{fig4}(b) respectively. The elliptic flow anisotropy for the clustered and unclustered cases is found to be close to each other which is consistent with the average initial elliptic eccentricities obtained for the respective cases. We see a slightly larger elliptic flow at $ p_T \approx 2$ GeV for the unclustered case in comparison to the clustered case.

However, $v_3$ for the clustered case is found to be twice as large as the same obtained for the unclustered case.  It is to be noted that the orientation averaged triangular flow parameter for the clustered case is still significantly large and similar to the $ v_3(p_T)$  obtained from most-central ($ b\approx0$ fm) collisions of $\alpha$-clustered carbon with Au nucleus at an orientation angle $ \theta=\pi/4$ (see Fig. 5(a) of Ref. ~\cite{Dasgupta:2020orj}) considering smooth initial density distribution. Additionally, such large triangular flow anisotropy is also comparable to the direct photon $v_3$ data obtained for $20-40\%$ Au+Au collisions at RHIC. It is to be noted that a larger initial thermalization time $\tau_0$ or a smaller freeze-out temperature ($T_f$) would further increase the value of thermal photon $ v_3$ as discussed in the Ref.~\cite{Dasgupta:2020orj}. These results clearly state that photon $v_3$ in relativistic nuclear collisions can efficiently reflect the initial state triangular anisotropy associated with the triangular $\alpha$-cluster structure in carbon nucleus.  It is also well known that a significant contribution of the prompt photons appears in the region $ p_T >3$ GeV~\cite{Dasgupta:2020orj} in the direct photon spectrum and it dominates over the thermal radiation in that $p_T$ range. However, the prompt photons do not contribute directly to the anisotropic flow.  These non-thermal photons only dilute the flow parameters in the larger $p_T$ region by adding extra weight in the denominator of Eq. 5. One can still expect to get a large direct photon $ v^{dir}_3(p_T)$  after including the non-thermal prompt contribution in the calculation~\cite{Dasgupta:2020orj}.\\
\begin{figure}[htbp!]
	\centering
	{\includegraphics*[scale=0.4,clip=true]{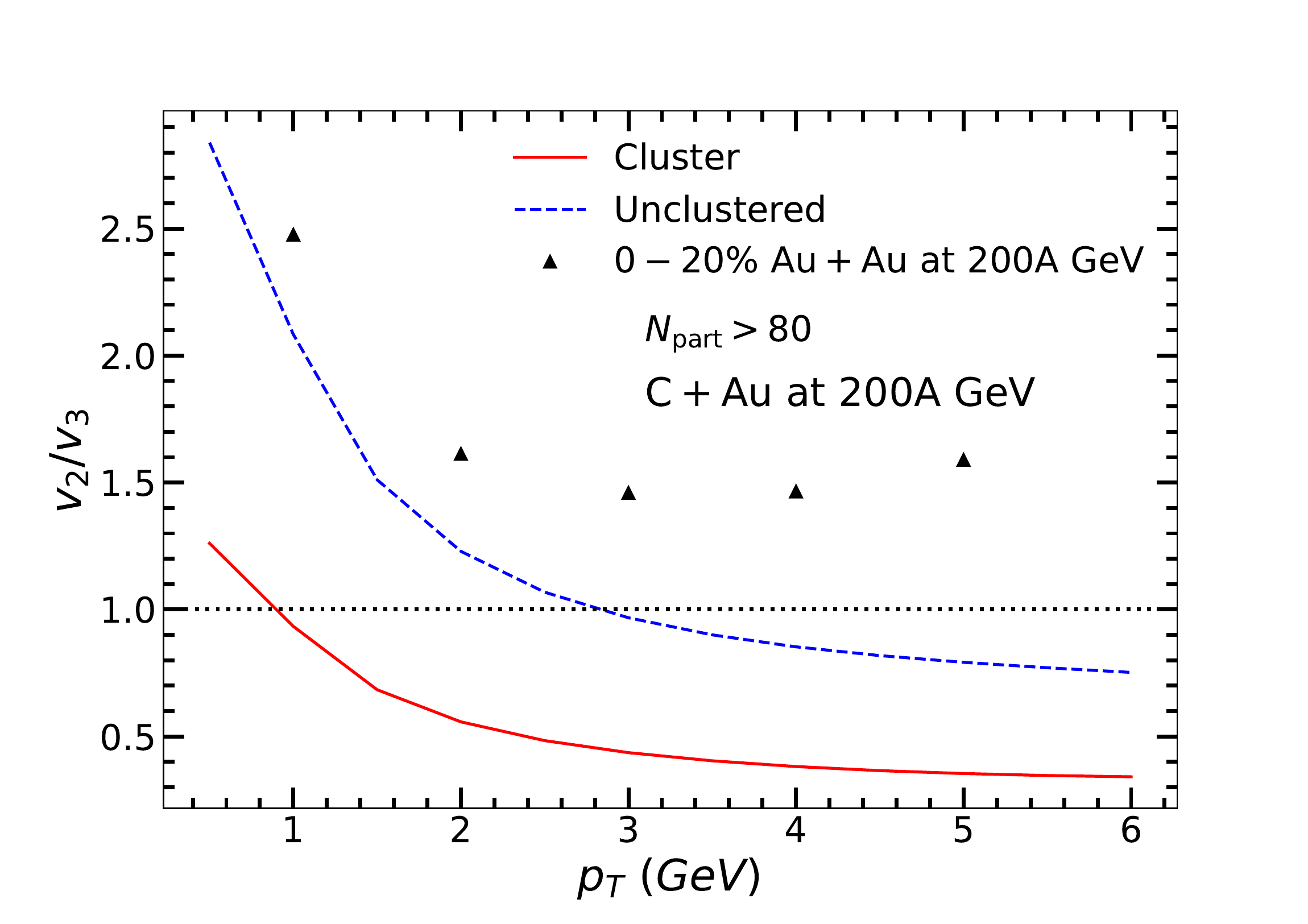}}
	\vspace{-3mm}
	\caption {(Color online) The ratio of thermal photon $v_2$ and $v_3$  as function of $ p_T$ for the $\alpha$-clustered and unclustered C+Au collisions at $200A$ GeV. The  ratio of thermal photon $v_2$ to $v_3$ for $0-20\%$ Au+Au collisions at RHIC using  the same hydrodynamic framework is shown for comparison~\cite{Adare:2015lcd}. }
	\label{fig5}
\end{figure}

In a recent study, it has been emphasized that the ratio of anisotropic flow coefficients ($ v_n/v_m$) can minimize the uncertainties arising due to the non-thermal  contributions~\cite{Chatterjee:2021bhz} in the photon anisotropic flow calculation. The ratio is shown to be a potential observable to probe the thermal phase contribution in the direct photon anisotropic flow. Although the individual flow parameters ($v_2$, $v_3$) are found to underestimate the experimental $v_n$ data { of Au+Au collisions} at RHIC, the photon $v_2/v_3$   {of the same} is found to be close to the PHENIX data in the $p_T$ region 2 -- 3.5 GeV, which is believed to be dominated by the thermal radiation. The ratio is also found to be sensitive to the initial conditions of the model calculation in different $p_T$ regions compared to the individual flow parameters. The ratio of directed flow parameter with photon $v_2$ (or $v_3$) (along with the individual photon anisotropic flow parameters) provides additional information to constrain the initial parameters of the model calculation. We show the ratio of thermal photon $ v_2$ to $ v_3$ as a function of $p_T$ in Fig.~\ref{fig5}. The dashed line shows the ratio for the unclustered case which is found to be about 2 at $p_T \approx 1$ GeV and above $p_T>3$ GeV, the ratio gets closer to 1. However, for the clustered case the ratio (solid line) is found to be smaller than 1 in the region $ p_T > 1$ GeV, which in turn indicates a significantly larger thermal photon $ v_3$  compared to the thermal photon $ v_2$. {We simultaneously show the ratio of thermal  photon $v_2$ and $v_3$ from $0-20\%$ Au+Au collisions at RHIC  (solid triangles) using the same hydrodynamical framework for comparison{~\cite{Chatterjee:2021bhz}}. We find that the ratio for the central Au+Au collisions is closer to the result from unclustered C+Au collisions whereas the ratio from the clustered case is significantly smaller in the thermal (2 -- 4 GeV) $p_T$ range. Thus, experimental determination of photon $v_2/v_3$ ratio from C+Au collisions can be an important observable to identify the clustered structure in carbon nucleus.  \\

\begin{figure}[htbp!]
	\centering
	{\includegraphics*[scale=0.4,clip=true]{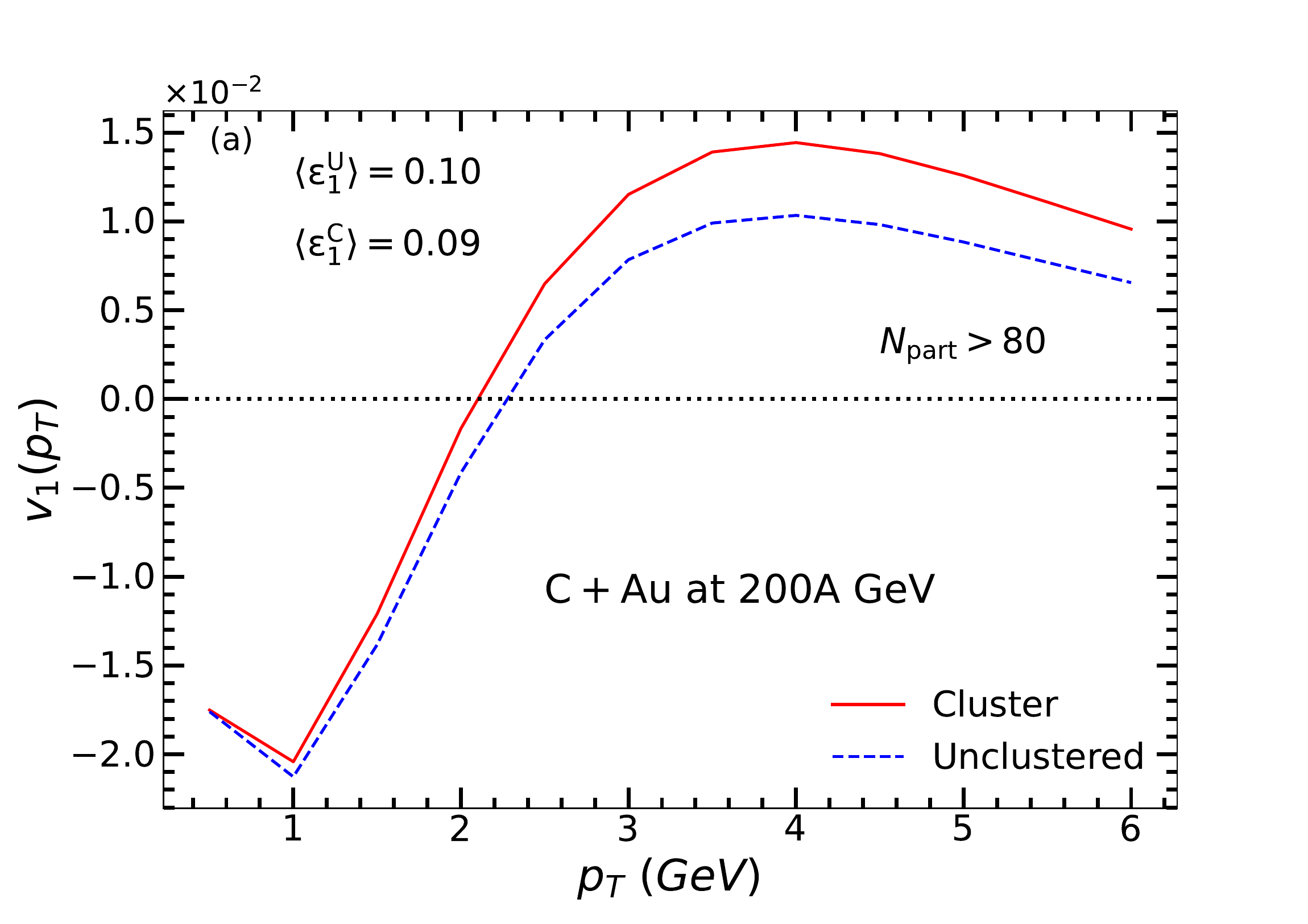}}
	{\includegraphics*[scale=0.4,clip=true]{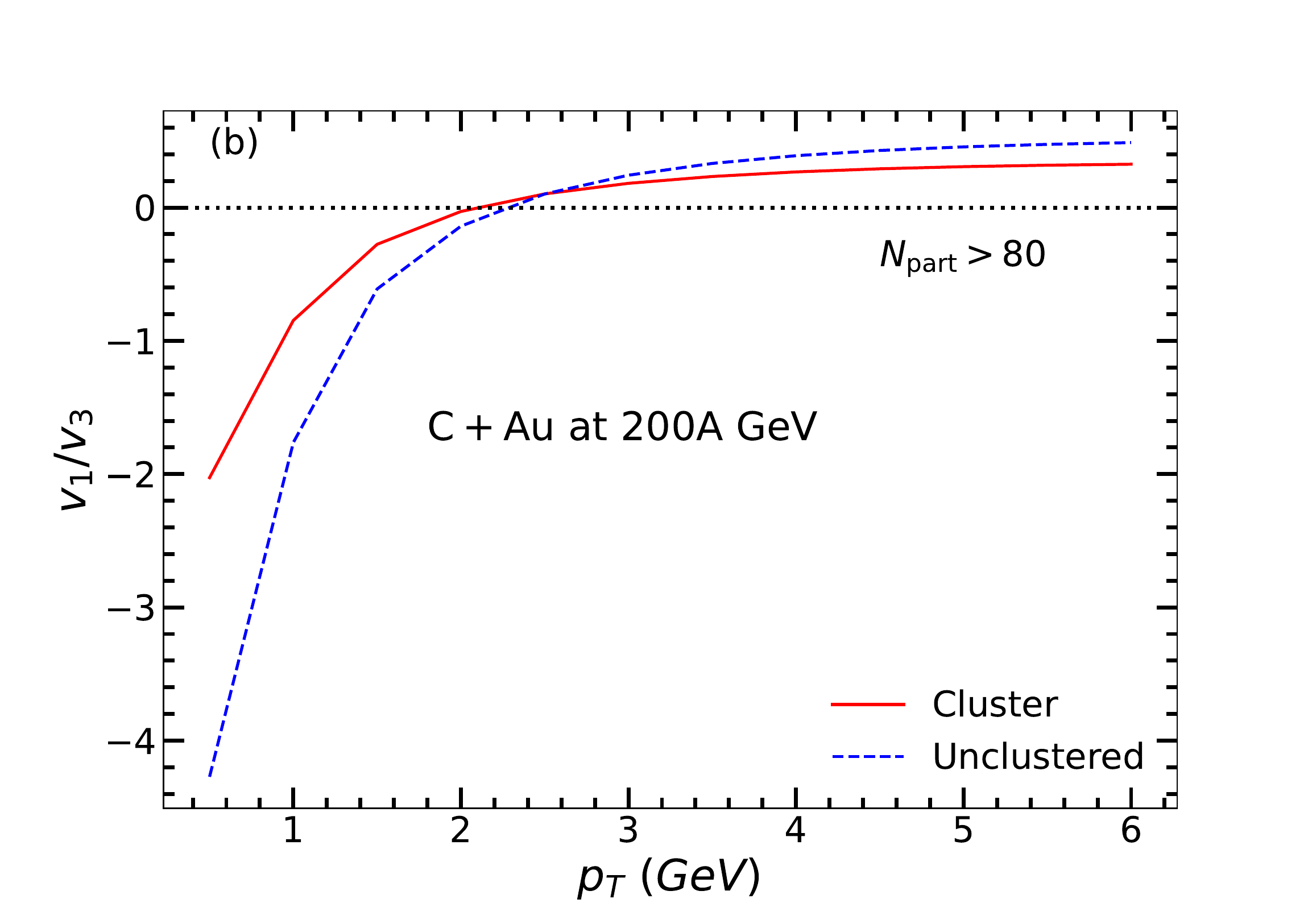}}
	\vspace{-3mm}
	\caption {(Color online)  (a) Directed flow of thermal photons and (b) its ratio with thermal photon $ v_3$  for the $\alpha$-clustered C+Au and the unclustered C+Au collisions at $200A$ GeV.}
	\label{fig6}
\end{figure}  
The directed flow of thermal photons from C+Au collisions is shown as a function of $ p_T$  in Fig.~\ref{fig6}(a). The photon $v_1$ for both clustered and unclustered cases is found to be close to each other. This observation is consistent with the directed initial eccentricities obtained for the two cases. The linear correlation coefficient between $p_T$-integrated photon $v_1$ and $\varepsilon_1$ is found to be about -0.59 and -0.51 for the clustered and the unclustered cases, respectively  {(see Fig.~2 in the Supplemental Material~\cite{supplemental})}. We show the ratio of the directed flow of thermal photons with triangular flow in Fig.~\ref{fig6}(b). The ratio is seen to be different in the region $ p_T < 2$ GeV where the ratio for the clustered case is found to be larger than the unclustered case. In the region $ p_T >2$ GeV an opposite behavior is observed, however, the difference between the two is relatively smaller. The $v_1/v_3$ has been shown to be more sensitive to the hadronic phase unlike  $v_2/v_3$ of photons~\cite{Chatterjee:2021bhz}. Thus, the ratio of photon $v_n$ can also be valuable to know about the freeze-out temperature of the evolving fireball. \\

Therefore, this study overall provides a qualitative understanding of thermal photon production and anisotropic flow from most-central $\alpha$-clustered C+Au collisions at $200A$ GeV. { It is to be noted that we have checked all the photon $v_n$ results with the scalar product method  ~\cite{David:2019wpt,Shen:2013cca} and also considering the lattice-based equation of state s95p-v1~\cite{Huovinen:2009yb} to see the sensitivity of our observations with the choice of the framework (see Appendix-II). The qualitative nature of the results is found to remain the same irrespective of the method we chose. We also notice that the difference between the results on a quantitative scale is marginal only. { A (3+1) dimensional viscous hydrodynamical calculation is ideal to obtain more reliable results for C+Au collisions on a quantitative scale; however, according to Fig. 4 in the ref. \cite{Shen:2016zpp}, we cannot expect a significant change in the results for the most-central collision scenario.}

\section{SUMMARY}
We calculate the production and anisotropic flow of thermal photons from most-central $\alpha$-clustered ${\rm C}$ and ${\rm Au}$ collisions at $200A$ GeV using an event-by-event hydrodynamic framework and compare the results with those obtained from unclustered carbon and gold collisions. The slope of the thermal photon spectrum from the clustered carbon is found to be similar to the same obtained from the unclustered carbon. However, the clustered structure affects the initial triangular eccentricity as well as the triangular flow parameter of photons significantly compared to the unclustered carbon and gold collisions.  Although we find a similar thermal photon $v_2$ for both clustered and unclustered carbon collisions, the thermal photon $v_3$ for the clustered case is found to be almost twice as large as the unclustered case. 

In addition, we show that the ratio of photon anisotropic flow parameters ($v_2/v_3$) can be a useful observable to recognize the clustered structure in carbon nucleus.  The  $v_2/v_3$ ratio is found to be strongly sensitive to the $\alpha$-cluster structure and is significantly suppressed compared to the unclustered case. We also show that the directed flow parameter of photons can be a potential observable along with the elliptic and triangular flow parameters to constrain the initial state produced in heavy ion collision. We conclude that photon flow observables are potential probes to detect the  $\alpha$-cluster structure in carbon nucleus and experimental determination of the anisotropic flow parameters from C+Au collisions can be useful to study the initial state in relativistic heavy ion collisions.

\begin{acknowledgments}
	PD very much appreciates the generous help from Dr. Han-Sheng Wang and thanks the Fudan cluster computing facilities at Fudan. RC thanks the Kanaad and GRID computer facilities at VECC. GM thanks Prof. Bo Zhou for the helpful discussions. This work is supported by the National Natural Science Foundation of China under Grants  No.12147101, No. 12150410303 , No. 11890714, No. 11835002, No. 11961131011, No. 11421505, the National Key Research and Development Program of China under Contract No. 2022YFA1604900, the Strategic Priority Research Program of Chinese Academy of Sciences under Grant No. XDB34030000, and the Guangdong Major Project of Basic and Applied Basic Research under Grant No. 2020B0301030008 (G. M. and P. D.).  
\end{acknowledgments}
{ \section*{Appendix-I}

In this section, we discuss in detail the initial conditions of our model calculations and the method for calculating thermal photon flow observables. \\

{\it Initial Conditions} : We consider a triangle-shaped $\alpha$-clustered carbon where the vertices of 3 $\alpha$-clusters reside at 3 corners of an equilateral triangle. The side length ($l$) of the triangle is $3.05$ fm  and the radius of each cluster ($r_\alpha$) is  $0.96$ fm. The nuclear density distribution corresponding to each cluster follows  :
	\begin{eqnarray}
		f_i(\vec{r})=A \exp \left (- \frac{3}{2} \, (\vec{r}-\vec{c_i})^2/r_\alpha^2 
		\right ), 
		\label{alpha_dist}
	\end{eqnarray}
	where, $\vec{c_i}$ denotes the position of the center of $i^{\rm th}$ cluster in carbon. For the unclustered case, we take a 2-parameter Wood-Saxon density profile in such a way that the root mean square radius of the unclustered carbon is similar to the clustered carbon ($\approx 2.26$ fm). 
	
	A two-component Monte Carlo Glauber (MCG) model framework is used to distribute the initial entropy density on the transverse  plane  for each event where entropy density at any transverse coordinate $(x,y)$ is obtained using the following relation :
	\begin{eqnarray}
		s(x,y)=K\sum_{i,j=1}^{N_{\rm part},N_{\rm coll}} [  \, \nu \, n_{\text{coll}}(x_i,y_i) F_{i}(x,y)
		\nonumber \\ +(1-\nu)\, n_{\text{part}}(x_j,y_j)\ F_{j}(x,y) ] \, .
		\label{imple_eqn}
	\end{eqnarray}
	In the above equation,  $n_{\rm coll}$ and $n_{\rm part}$  denote the number of  binary collision and participant sources at the $(x_i,y_i)$ and $(x_j,y_j)$ positions, respectively. The  values of nucleon-nucleon inelastic cross section ($\rm \sigma_{NN}$),  hardness factor ($\nu$) and the normalization constant ($K$) are taken as $42$ mb, $0.145$ and $81$ fm$^{-2}$, respectively. The function $F_{i}(x,y)$ [or $F_{j}(x,y)$] is a normalized Gaussian distribution centering about the $i^{th}$ collision (or $j^{th}$ participant) source, 
		\begin{equation}
			\label{normal_dist_eq}
			F_{i,j}(x,y)=\frac{1}{2\pi \sigma ^2}\, e^{ -\frac{(x-x_{i,j})^2+(y-y_{i,j})^2}{2\sigma ^2}},
		\end{equation}
		where the Gaussian smearing width ($\sigma$) of radius around each collision and participant source is considered as $0.4$ fm~\cite{Dasgupta:2020orj}. \\
	 The above set of the initial hydrodynamic parameters has been found to satisfactorily explain the charged hadron multiplicity and differential spectrum of $\pi^0$ at the mid rapidity for most central (i.e., 0 - 5\%) $^3$He+Au collisions at $200A$ GeV at RHIC~\cite{PHENIX:2021dod} 	(see Fig.1 in the Supplemental Material~\cite{supplemental}).
		
	{\it Participant Plane Method} : The differential anisotropic flow coefficients ($v_n(p_T)$ for  $n=1,2,$ and $3$) for each event are obtained as :
	
	\begin{eqnarray}
		v_n (p_T) \ = \ \frac{\int_0^{2\pi} \, 
			d\phi\,{\rm cos} [\,n (\phi-\psi_{n})]\,\frac{dN}{p_T dp_T dy d\phi}}{\int_0^{2\pi}\,d\phi\,\frac{dN}{p_Tdp_Tdy d\phi}},
		\label{v2flow}
	\end{eqnarray}

	where,  $\phi$ is the azimuthal angle of particle's momentum and $\psi_n$ is the participant plane angle. 
	We calculate the event-averaged final flow observables ($\langle v_n\rangle$, which we denote hereafter as $v_n$ for simplicity) by using the following equation,
	\begin{eqnarray}
		&\langle v_n(p_T)\rangle =\frac {\sum_{i=1}^{\rm N_{events}}\frac{dN^{(i)}}{d^2p_Tdy} v_n^{(i)}(p_T)}{\sum_{i=1}^{\rm N_{events}}\frac{dN^{(i)}}{d^2p_Tdy}}
		\label{vnavg}
	\end{eqnarray}
	In the above equations, the superscript `$i$' corresponds to the $i$th  event.
	The participant plane angle is determined by,
	\begin{equation}
		\psi_{n} = \frac{1}{n} \arctan 
		\frac{\int \mathrm{d}x \mathrm{d}y \; r^2 \sin \left( n\Phi \right) \epsilon\left( x,y,\tau _{0}\right) } 
		{ \int \mathrm{d}x \mathrm{d}y \; r^2 \cos \left( n\Phi \right) \epsilon\left( x,y,\tau _{0}\right)}  + \pi/n \, ,
		\label{sai_nm}
	\end{equation}
	The initial-state eccentricities ($\rm \varepsilon_n$ where $n=1,2,$ and $3$)  for each event are obtained using the  relation,
	\begin{eqnarray}
		\varepsilon_{n}=-\frac{\int \mathrm{d}x \mathrm{d}y \; r^2 \cos [\,n (\Phi-\psi_n)]    \epsilon\left( x,y,\tau _{0}\right) } 
		{ \int \mathrm{d}x \mathrm{d}y \; r^2  \epsilon\left( x,y,\tau _{0}\right)},
		\label{e_mn1}
	\end{eqnarray}
	where $\Phi$ and $r$ are spatial azimuthal angle and the radial distance, and $\epsilon$ is the energy density on the transverse plane. 
	}
{\section*{Appendix-II}
	%	\vspace{0.25cm}
	\begin{figure}[htbp!]
		\centering
		{\includegraphics*[scale=0.4,clip=true]{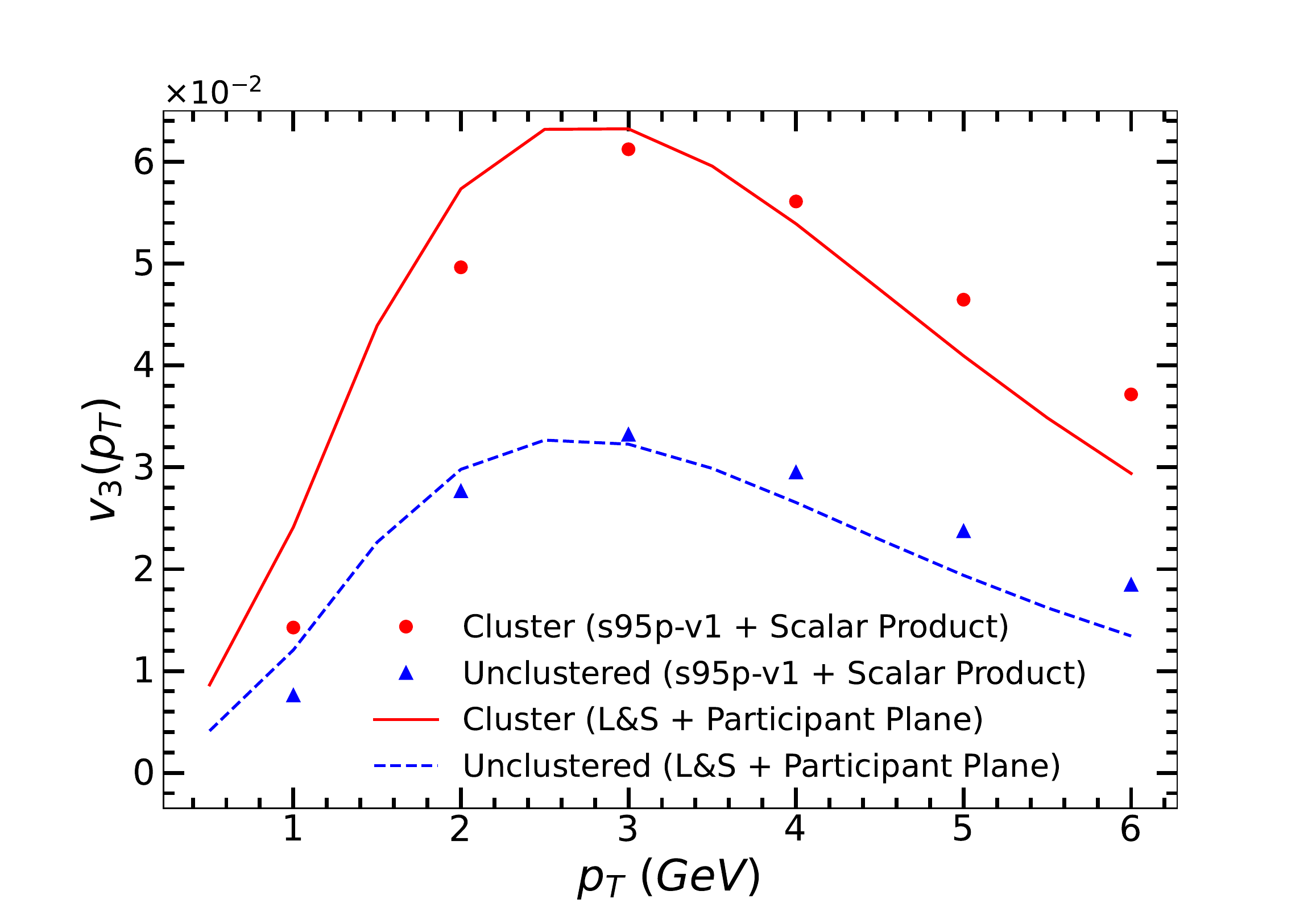}}
		{\includegraphics*[scale=0.4,clip=true]{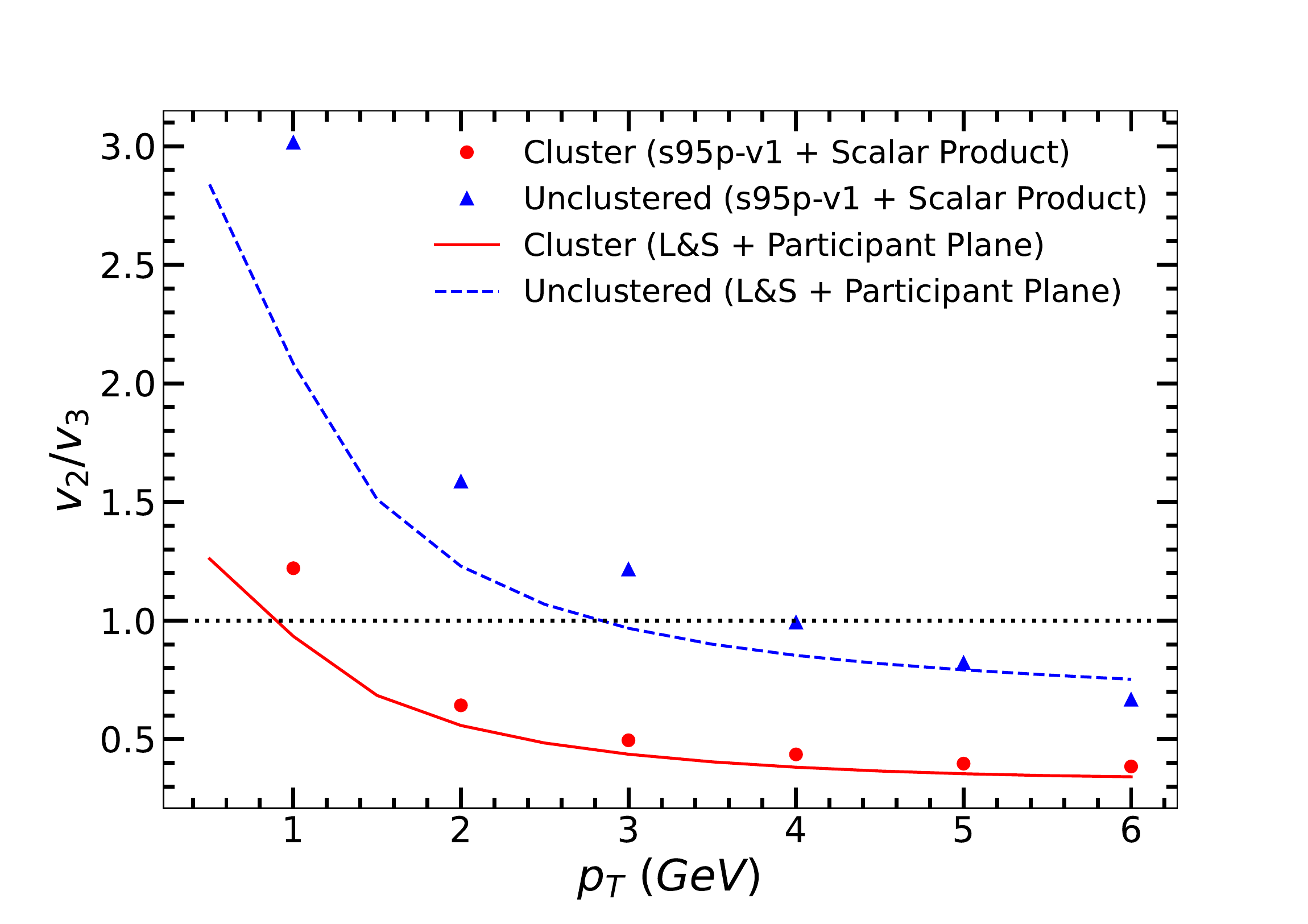}}
		\vspace{-3mm}
		\caption {(Color online)  The thermal photon (a) $v_3$ and (b) the ratio  $v_2/v_3$ as a function of $p_T$ using the scalar product method in the MUSIC ideal hydrodynamical framework for the $\alpha$-clustered C+Au and the unclustered C+Au collisions at $200A$ GeV. The results have been compared with results presented earlier in Fig.~\ref{fig4}(b) and Fig.\ref{fig5}. }
		\label{fig7}
	\end{figure} 
	{{\it Scalar Product Method}}: In this section, we calculate the photon flow observables from the most-central  C+Au collisions at $200A$ GeV using the scalar product method~\cite{David:2019wpt,Shen:2013cca}. Unlike the participant plane method, the scalar product method directly correlates the photon emission with the charged hadron event-plane angle. Therefore, it is more relatable to the experimental data.
We use  Monte-Carlo Glauber initial conditions  (Eq.\ref{imple_eqn})  along with lattice-based equation state s95p-v1~\cite{Huovinen:2009yb} within the MUSIC ideal hydrodynamical framework~\cite{Schenke:2010nt} to calculate the spacetime evolution of the fireball. The other associated hydrodynamic parameters have been considered to be the same as presented in Section II.  \\

In Fig.~\ref{fig7}(a), we show the thermal photon $v_3$ as a function of $p_T$ for the clustered and unclustered C+Au collisions using the scalar product method. The obtained results (represented by the solid symbols) are found to be very close to the earlier results using the participant plane angle method (represented by lines). Similar behavior has been found for the photon $v_2/v_3$ ratio as a function of $p_T$, as shown in Fig.~\ref{fig7}(b). We conclude that the calculation method merely affects the qualitative difference between the photon flow observables from the clustered and unclustered C+Au collisions. }

\end{document}